\documentclass[reprint,superscriptaddress,nofootinbib,floatfix]{revtex4-2}

\usepackage{graphicx}
\usepackage{amsmath}
\usepackage{amssymb}

\graphicspath{{figs/}}
\newcommand{\Ohn}{\mathit{Oh}}
\newcommand{\Bo}{\mathit{Bo}}
\newcommand{\La}{\mathit{La}}

\begin{document}

\title{The ejected volume and drop statistics by bubble bursting in a sessile drop}

\author{Alfonso M. Ga\~n\'an-Calvo}
\email{amgc@us.es}
\affiliation{Depto.\ de Ingenier\'{\i}a Aeroespacial y Mec\'anica de Fluidos,
Escuela T\'ecnica Superior de Ingenier\'{\i}a, Universidad de Sevilla,
41092 Sevilla, Spain}

\begin{abstract}
A bubble bursting inside a drop resting on a substrate can eject jet droplets
where a bubble at a bath would not. We compute how much liquid it ejects and
how the ejected droplets are distributed in size. The volume ejected by the jet grows by
more than an order of magnitude as the Bond number decreases from $0.5$ to
$0.1$ and then saturates, and the larger the Bond number, the smaller the
viscosity that suppresses ejection. At zero Bond number the substrate alone
extends the ejection boundary to larger Ohnesorge numbers, contrary to
expectation, and yet reduces the volume ejected at the same distance from it. Gravity
sets how much liquid the jet carries, not how the jet fragments: scaled with
its mean, the droplet-size distribution is independent of the Bond number and
has an exponential tail, as for free hollow droplets. The simulations, built on
a closed family of sessile equilibria, reproduce the measured velocity of the
first droplet within 7\,\%, while the volumes emitted in experiments are eight
to ten times those computed for the same geometry.
\end{abstract}

\maketitle

\section{Introduction}\label{sec:intro}

Bubbles bursting at the surface of a liquid eject jet droplets that are a major
source of marine aerosol and that carry dissolved and suspended material into
the air \citep{Spiel1994,DuchEtal2002,LhuiVill2012,Veron2015,GananCalvo2017,
Deike2018,DeikeReichlPaulot2022}. The canonical configuration is a bubble at
the flat surface of an unbounded bath. Real bursting events, however, often
happen in a bounded volume of liquid: in the drops and ligaments of a
breaking wave, in films and puddles, or in drops resting on a solid. The
companion study of free hollow droplets \citep{HollowDrops2026} shows that
confinement extends the ejection of droplets to bubbles too small to eject at a
flat surface, and that the extent of this extension, the ejected volume and the
droplet census are organized by a single geometric parameter.

A drop resting on a substrate is the most accessible confined geometry in the
laboratory. Experiments carried out concurrently with the present study by a
team led by the group that leads the partner subproject of the same coordinated
research project
\citep{GutierrezEtal2026} show that a
bubble bursting in a sessile drop can eject droplets below the threshold
Laplace number of a bath. Those experiments provide the measurements against
which the computations of this paper are verified.

Three aspects of the sessile problem are not settled by the existing work, and
they are the subject of this paper.

The first is the initial state. A sessile drop with a bubble at its apex is
described by four parameters (the Bond and Laplace numbers, the footprint
$R_D$ and the apex height $H$, or the liquid volume), and its equilibrium is
usually computed case by case. We show in \S\,\ref{sec:eq} that the whole
family follows from the classical bath solution by one vertical displacement,
and that the natural parameter is the overpressure of the drop. The second is the edge of the ruptured film.
Simulations of bubble bursting start from the cavity left by the film, and the
sharp edge is usually rounded by a small torus whose radius is chosen
\citep{DuchEtal2002}: a regularization parameter that stands for the
unresolved film thickness and for the instant at which the clock starts. In \S\,\ref{sec:rim}
the rim is built from the equilibrium itself, conserving the volume of the
liquid it replaces and moving at the Taylor--Culick speed
\citep{Taylor1959,Culick1960}, so that no free parameter remains. The third is
what the bursting ejects: how much liquid, in how many droplets, and with what
size distribution, as a function of gravity and viscosity at fixed geometry.
These are the quantities that enter aerosol source functions
\citep{Berny2021,DeikeReichlPaulot2022}, and they have not been computed for
the sessile geometry. Finally, the family of equilibria contains the free
hollow droplet, cut by the substrate, as its limit of zero Bond number. This
isolates the effect of the substrate on the ejection boundary, whose sign turns
out to be the opposite of the one expected.

The paper is organized as follows. The formulation, the initial condition and
the numerical method are given in \S\,\ref{sec:form}, and the method is
verified against experiment in \S\,\ref{sec:val}. Section~\ref{sec:sweep} gives the ejected volume over a
sweep in Bond and Ohnesorge numbers at fixed drop footprint and volume, and
\S\,\ref{sec:sizes} the droplet-size statistics. Section~\ref{sec:floor}
isolates the effect of the substrate at zero Bond number. In \S\,\ref{sec:Ve} the
computed ejected volume is set against the measured one. Conclusions follow in \S\,\ref{sec:concl}.

\section{Formulation}\label{sec:form}

\subsection{Configuration and parameters}

A gas bubble of volume $4\pi R_b^3/3$ rests at the apex of a liquid drop of
density $\rho$, viscosity $\mu$ and surface tension $\sigma$, which sits on a
horizontal substrate with its contact line pinned at radius $R_D$. The liquid is Newtonian and clean, so that $\sigma$
is uniform. Lengths are
scaled with $R_b$, velocities with $(\sigma/\rho R_b)^{1/2}$, times with
$t_\sigma=(\rho R_b^3/\sigma)^{1/2}$ and pressures with $\sigma/R_b$. The
parameters are the Bond and Ohnesorge numbers,
\begin{equation}
\Bo=\frac{\rho g R_b^2}{\sigma},\qquad \Ohn=\frac{\mu}{(\rho\sigma R_b)^{1/2}}=\La^{-1/2},
\label{eq:params}
\end{equation}
the footprint $R_D$ and the ratio of liquid to bubble volume,
$\Lambda=V_l/V_b$; the apex height $H$ and the apparent contact angle $\alpha$
follow from the equilibrium. The viscocapillary length is
$\ell_\mu=\mu^2/\rho\sigma=\Ohn^2R_b$ and the viscocapillary velocity
$v_\mu=\sigma/\mu=\Ohn^{-1}(\sigma/\rho R_b)^{1/2}$.

\subsection{The initial equilibrium}\label{sec:eq}

The axial coordinate $z$ is measured from the hydrostatic zero, so that the
liquid pressure is $p_l=p_a-\Bo\,z$ in scaled form. Along each meridian, with
arclength $s$ and tangent angle $\theta$, the bubble wall and the outer
meniscus satisfy the Young--Laplace equations
\begin{equation}
\begin{gathered}
\theta'+\frac{\sin\theta}{r}-\Bo\,z-P=0\quad\text{(bubble)},\\
\theta'+\frac{\sin\theta}{r}-\Bo\,z=0\quad\text{(meniscus)},
\end{gathered}
\label{eq:YL}
\end{equation}
with $r'=\cos\theta$ and $z'=\sin\theta$, where $P$ is the overpressure of the
gas over the atmosphere at $z=0$. The two arcs meet at the edge of the
spherical cap of the film, whose radius is $R_c=4/P$ since the film has two
interfaces. The unknowns, the height of the bubble bottom and $P$, are
fixed by the bubble volume $V_b=4\pi/3$ and by the height of the junction.

At a bath the junction lies at a height $z_0(\Bo)$ fixed by the classical
solution \citep{Toba1959,LhuiVill2012}, which is universal: for each $\Bo$ it
gives the junction radius, height and angle, and the arclengths of both arcs.
The observation on which the construction rests is that a sessile drop is the
same local problem displaced by $\Delta z<0$ with respect to the hydrostatic
zero. The meniscus is then continued from the displaced junction down to the
substrate, and the liquid volume $\Lambda$ fixes where the substrate cuts it,
and with it $R_D$ and $\alpha$. The cycle of Laplace jumps closes only if the
bubble pressure absorbs the hydrostatic offset, and the drop then carries an
overpressure
\begin{equation}
\Pi=-\Bo\,\Delta z,
\label{eq:Pi}
\end{equation}
in units of $\sigma/R_b$. This is the capillary pressure of the curved drop
acting on the bubble, the first of the confinement mechanisms discussed for free
hollow droplets by \citet{HollowDrops2026} (see also \citealt{GutierrezEtal2026}). A consistent formulation requires that
the hydrostatic reference of the outer meniscus move with the bubble; if only
the bubble is displaced, the loop of pressure jumps fails to close by
$-\Bo\,\Delta z$. With $\Delta z=0$ the solution reproduces the bath to
$10^{-6}$.

\begin{figure*}
\centering
\includegraphics[width=\textwidth]{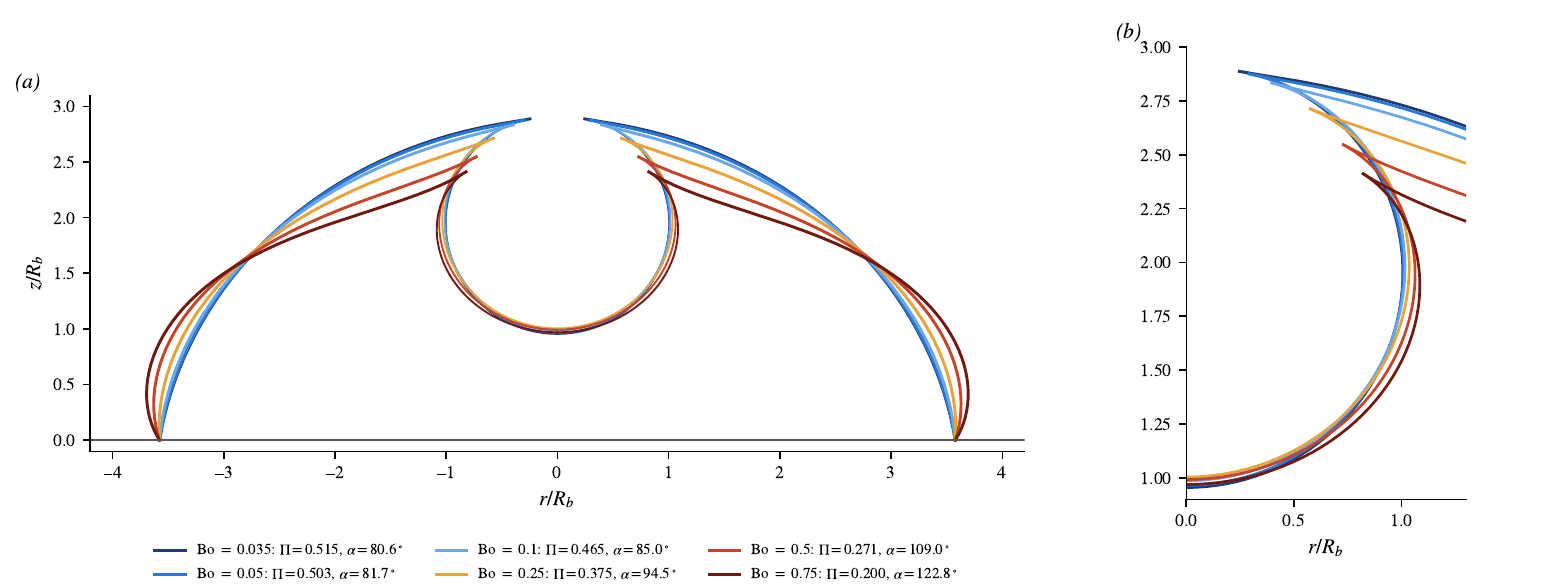}
\caption{Initial equilibria at fixed footprint $R_D/R_b=3.577$ and volume ratio
$\Lambda=16$. (\textit{a}) Bubble and meniscus for $\Bo$ from $0.035$ to
$0.75$; for each $\Bo$ the displacement $\Delta z$, hence the overpressure
$\Pi$, is solved so that $R_D$ is the same, and the contact angle $\alpha$
adjusts. The case $\Bo=0.75$ is shown for the geometry only. (\textit{b})
Detail of the bubble and of the meniscus near the film edge. The film, removed
at $t=0$, is not drawn.}
\label{fig:eq}
\end{figure*}

Each pair $(\Bo,\Pi)$ thus defines a one-parameter set of drops, indexed by
$\Lambda$ (or $R_D$, or $H$), bounded above by $\alpha=180^\circ$ and below by
$\Lambda_{\min}(\Bo,\Pi)$, at which the substrate reaches the bottom of the
bubble: $z_b=0$, with $z_b$ the height of the bubble bottom above the
substrate. This is the frontier of the problem considered: for a given amount
of liquid it sets the largest bubble, $V_b=V_l/\Lambda_{\min}$, beyond which the
bubble rests on the substrate with a contact line of its own. Near it the
substrate also acts on the flow beneath the cavity, an effect isolated in
\S\,\ref{sec:floor}. A prescription of $R_D$, $V_l$
and $V_b$, as in the experiments, is recovered by a one-dimensional search in
$\Pi$. We tabulated the family for 81 values of $\Bo$ between $10^{-4}$ and
$70$ and $0\le\Pi\le2$ in steps of $0.025$ (3\,458 families; supplementary
material). Figure~\ref{fig:eq} shows the equilibria at fixed footprint
$R_D=3.577$ and volume ratio $\Lambda=16$ used in \S\,\ref{sec:sweep}: as $\Bo$
increases at fixed footprint, $\Pi$ decreases, the contact angle opens and the
meniscus between the film edge and the shoulder of the drop develops an
inflection imprinted by the bubble.

\paragraph{Small Bond number.} At a bath, the small-Bond-number limit was
analysed by \citet[appendix B]{Howell1999}: to leading order the bubble is a
sphere under a flat surface, and the outer meniscus must be matched to the far
field, as for a meniscus on a small cylinder. In a sessile drop whose footprint
is of the order of the bubble there is no far field, and the limit at fixed
$\Pi$ is regular. The displacement diverges, $\Delta z=-\Pi/\Bo$, but
$\Bo\,z=-\Pi+O(\Bo\,H)$ over the whole drop, so that both equations
(\ref{eq:YL}) reduce, as expected, to surfaces of constant mean curvature. The
table shows which. At $\Bo=10^{-4}$, $P=2+\Pi$ to within $10^{-4}$ for every
$\Pi$ between 0 and 2, and the junction radius is $0.012$--$0.05$, of order
$\Bo^{1/2}$ as at a bath: the film shrinks to a point at the apex, the bubble
is a unit sphere, and the meniscus, which leaves the junction tangent to it, is
a sphere of radius $2/\Pi$. The small-Bond-number limit of the sessile drop is
therefore the hollow drop of \citet{HollowDrops2026}, a bubble internally
tangent to a spherical drop of radius ratio $\lambda=2/\Pi$, cut by the
substrate; as $\Pi\to0$, $\lambda\to\infty$ and the bath is recovered. The
bounds of the family follow in closed form: $\Lambda_{\min}=3\lambda-3=6/\Pi-3$,
and $\alpha=180^\circ$ at $\Lambda=\lambda^3-1$, which the table reproduces to
better than 1\,\% for $0.25\le\Pi\le1.9$. The limit requires the drop, not
only the bubble, to be small compared with the capillary length,
$\Bo\,\lambda^2\ll1$.

\subsection{The film edge: a parameter-free Taylor--Culick rim}\label{sec:rim}

In the equilibrium of \S\,\ref{sec:eq} the film is a membrane of zero
thickness and tension $2\sigma$, the usual limit for a film thickness
$h_f\ll1$, and, its tension $2\sigma$ balancing those of the bubble wall and
the meniscus, the three surfaces are tangent at the junction radius $r_j$. At $t=0$ the liquid thinner than a thickness $h_r$
is replaced by a rim of equal volume, the rule used for the hollow drop
\citep{HollowDrops2026}. There the film is the neighbourhood of a point of
tangency, and its first rupture lies necessarily on the axis. Here the film is
a cap of finite area: its thickness at rupture is set by drainage, which we do
not compute, and rupture nucleates mostly off the axis, at the foot of the cap
\citep{LhuiVill2012,PoulainEtal2018}. The axisymmetric initial condition does not depend on
where it nucleates provided $h_f\ll1$: a hole opening anywhere sweeps a cap of
arc length $O(1)$ at the Taylor--Culick speed $(2/h_f)^{1/2}$, in a time
$O(h_f^{1/2})$. In the water case of \S\,\ref{sec:GHA} ($R_b\approx1$\,mm) the
cap measures $0.99$ from edge to edge, and for $h_f=1\,\mu$m it is swept in
$0.02\,t_\sigma$, against $t\approx0.5$ to the first detachment. The cap is
therefore removed at once, and the liquid it carries, a fraction $O(h_f)$ of
the cavity volume, is neglected.

What remains is a cusp at $r_j$, beyond which the liquid between the bubble
wall and the meniscus forms a thin wedge whose thickness $h(r)$ grows
quadratically from zero, as the gap between the two spheres of the hollow
drop. Physically, the retracting film has collected its liquid into a
rim that moves at the Taylor--Culick speed $(2\sigma/\rho h)^{1/2}$
\citep{Taylor1959,Culick1960}. We build that rim from the equilibrium as
follows.

The wedge between $r_j$ and the radius $R_p$ at which its thickness reaches a
value $h_r$ is removed and replaced by a torus of tube radius $b$, centred at
mid-thickness at $r=R_p$. The radius $b$ is not chosen: it is found by
bisection so that the volume of the torus, after it has been blended into both
arcs with a smooth maximum over $0.5\,b$, equals the volume
$2\pi\int_{r_j}^{R_p}r\,h(r)\,\mathrm{d}r$ of the wedge it replaces. The liquid
of the rim is launched at the Taylor--Culick speed $(2/h_r)^{1/2}$ along the
mean tangent of the two arcs at $R_p$. The only length left, $h_r$, is set by
the resolution: $h_r=\max(2.5\Delta,h_r^{(3\Delta)})$, where $\Delta$ is the
smallest cell size and $h_r^{(3\Delta)}$ is the thickness for which the
volume-conserving tube radius is $3\Delta$, so that the rim is always resolved
by at least three cells. Mass is conserved to $10^{-6}$ (relative) at
initialization, and the rim carries the momentum the film would have given it.

A torus of prescribed radius started from rest adds instead a free parameter
and no momentum. Section~\ref{sec:GHA} shows, for a measured configuration,
that replacing our rim by
such a torus ($R_{\rm rim}=0.024$, from rest) leaves the collapse and the first
detachment unchanged and changes the ejected volume by less than 20\,\%. The
construction is therefore not needed to obtain the collapse right, but it
removes an arbitrary choice from the computation.

\subsection{Numerical method and droplet census}\label{sec:num}

The axisymmetric, incompressible two-phase Navier--Stokes equations with
surface tension and gravity are solved with Basilisk \citep{Popinet2009}, with
a geometric volume-of-fluid method, height-function curvature, balanced-force
surface tension and gravity in reduced form. The density and viscosity ratios
are $\rho_r=10^{-3}$ and $\mu_r=10^{-2}$. The substrate is a no-slip wall on
which the contact line is pinned by prescribing the volume fraction ($f=1$ for
$r<R_D$). The square domain has side $L_0=25.71$ (ten times the length
$(1+\Lambda)^{1/3}$ that scales the confined collapse at $\Lambda=16$; see
\citealt{HollowDrops2026}), so that droplets can be followed far from the drop.
At refinement level $N$ the smallest cell is $\Delta=L_0/2^N$: $R_b/319$ at
$N=13$, used in the sweep of \S\,\ref{sec:sweep}, and $R_b/637$ at $N=14$, used
in \S\,\ref{sec:val}.

The mesh is adapted by wavelet estimation of the error. We use two criteria.
The first refines on the volume fraction and on the two velocity components,
with tolerances $10^{-3}$, $10^{-2}$ and $10^{-2}$; we denote it
$\{f,\mathbf{u}\}$. The second adds the interface curvature $\kappa$ with tolerance $10^{-8}$, which drives the mesh to
the finest level along the whole interface; we denote it
$\{f,\mathbf{u},\kappa\}$ (the criterion used in the sessile-drop simulations
of \citealt{GutierrezEtal2026}). All the results of \S\S\,\ref{sec:sweep} and
\ref{sec:sizes} were computed with both. The choice is of little consequence
for what this paper measures: it changes the volume ejected during the jet
phase by about a quarter on average and leaves the scaled size distribution
unchanged; only the few late droplets of \S\,\ref{sec:late}, which are
sensitive to any perturbation, differ by more.

Every fragment of liquid (connected component with $f>1/2$) is tracked and
linked in time. A droplet \emph{escapes} if its radius is $R_d\ge3\Delta$, it
travels $1.5(1+\Lambda)^{1/3}$ beyond the apex of the drop with sustained
positive velocity, and it is not the mother drop. Escaped droplets are counted
at birth, and a droplet is \emph{persistent} if its track survives to the end
of the run or leaves the domain. Because every droplet travels on the axis, a
faster droplet that catches a slower one must coalesce with it; the difference
between the two counts measures this axisymmetric artefact
(\S\,\ref{sec:coal}). The ejected volume $M_e$ is the total volume of the
persistent escaped droplets, and $V_{\rm jet}$ the volume of the escaped
droplets born at $t\le1$ (in units of $t_\sigma$), which isolates the jet phase.
Runs stop automatically $0.5$ after the last birth.

\section{Verification against experiment}\label{sec:val}

The method is verified on two configurations, built with the construction of
\S\,\ref{sec:eq} from the apex height, footprint and Bond number and computed
at $N=14$: a viscous drop inside the range where ejection below the bath
threshold has been observed, and a water drop for which the first droplet and
the total emitted volume have been measured \citep{GutierrezEtal2026}.

\subsection{Ejection below the bath threshold}\label{sec:GHB}

\begin{figure*}
\centering
\includegraphics[width=\textwidth]{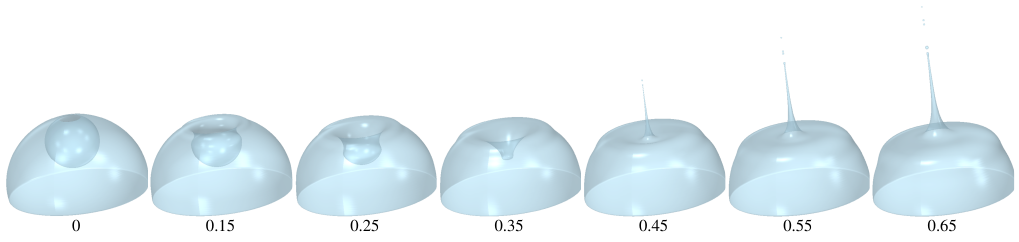}
\caption{Ejection below the bath threshold: $H/R_b=2.96$, $R_D/R_b=2.57$,
$\Bo=0.048$, $\La=468$, computed at $N=14$ ($\Delta=R_b/637$) with the initial
condition of \S\,\ref{sec:form}. The axisymmetric interfaces are drawn as
surfaces of revolution; the base of each frame is the area wetted on the
substrate. Times in units of $t_\sigma$.}
\label{fig:ghb}
\end{figure*}

At a bath no droplet is ejected below $\La\approx540$
\citep{WallsEtal2015,GananCalvo2017,Deike2018}; in sessile drops, ejection was
observed experimentally down to $\La\approx300$ \citep{GutierrezEtal2026}. We computed a viscous configuration inside that
range: $H=2.96$, $R_D=2.57$, $\Bo=0.048$ and $\La=468$ ($\Ohn=0.0462$), which
corresponds to $\Pi=0.725$, $\alpha=101.3^\circ$ and $\Lambda=9.55$
(figure~\ref{fig:ghb}). The surface inverts at $t\simeq0.35$--$0.40$, with the
bottom of the cavity $0.35R_b$ above the substrate, and the first droplet
detaches at $t=0.45$.
The computation ejects below the bath threshold, as the experiments do, and
quantifies the ejection: six droplets escape, with a total volume of
$1.17\times10^{-2}R_b^3$; the first has $R_d=0.028R_b=13\,\ell_\mu$ and leaves
at $v_d=19.8(\sigma/\rho R_b)^{1/2}=0.91\,v_\mu$.

\subsection{The first droplet in water}\label{sec:GHA}

\begin{figure*}
\centering
\includegraphics[width=\textwidth]{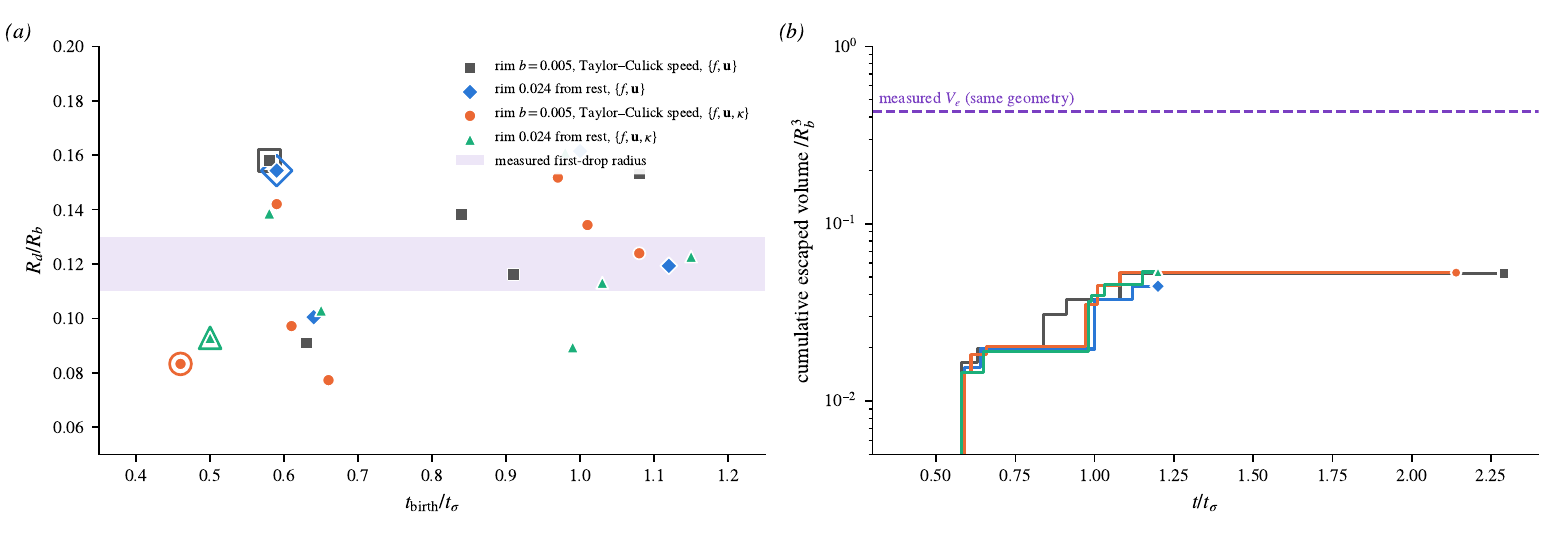}
\caption{Water drop with $H/R_b=3.94$, $R_D/R_b=1.99$, $\Bo=0.143$,
$\La=6.94\times10^4$, $N=14$, with two initial conditions and two refinement
criteria. (\textit{a}) Escaped droplets: radius against time of birth; the large
open symbol marks the first droplet (the one leading the train). Band: radius of
the measured first droplet. (\textit{b}) Cumulative escaped volume; the two
runs with our rim were continued until the automatic stop. Dashed line: total
volume emitted in the experiments at the same confinement, $V_e\simeq0.054R_D^3$
\citep{GutierrezEtal2026}.}
\label{fig:gha}
\end{figure*}

The second configuration is a water drop with $H=3.94$, $R_D=1.99$,
$\Bo=0.143$ and $\La=6.94\times10^4$ ($\Ohn=0.0038$). It corresponds to $\Pi=0.590$,
$\alpha=160.8^\circ$ and $\Lambda=16.04$. We computed it with four set-ups,
crossing two initial conditions (our rim, $b\simeq0.005$ at the
Taylor--Culick speed, and a torus of radius $0.024$ started from rest) with the
two refinement criteria (figure~\ref{fig:gha}).

The collapse is insensitive to both choices: neither the rim nor the
refinement criterion changes the inversion, and the first droplet detaches at
$t=0.46$--$0.59$ in all four set-ups. Its measured velocity,
$\approx13(\sigma/\rho R_b)^{1/2}$, is reproduced within 7\,\% by all four
($12.3$--$13.9$), and its measured radius, $\approx0.12R_b$, lies between the
values given by the two refinement criteria ($0.08$--$0.16$).

The ejected volume is also robust. It is $0.044$--$0.053R_b^3$
with every set-up, and the two runs continued until the automatic stop
($t=2.29$ and $2.14$) eject nothing after $t=1.1$: the fragments born later fall
back into the drop. The number of escaped droplets changes with the criterion
(five against seven), the volume does not. Its comparison with the measured
emitted volume is the subject of \S\,\ref{sec:Ve}.

\section{Ejected volume at fixed drop footprint}\label{sec:sweep}

\begin{figure*}
\centering
\includegraphics[width=\textwidth]{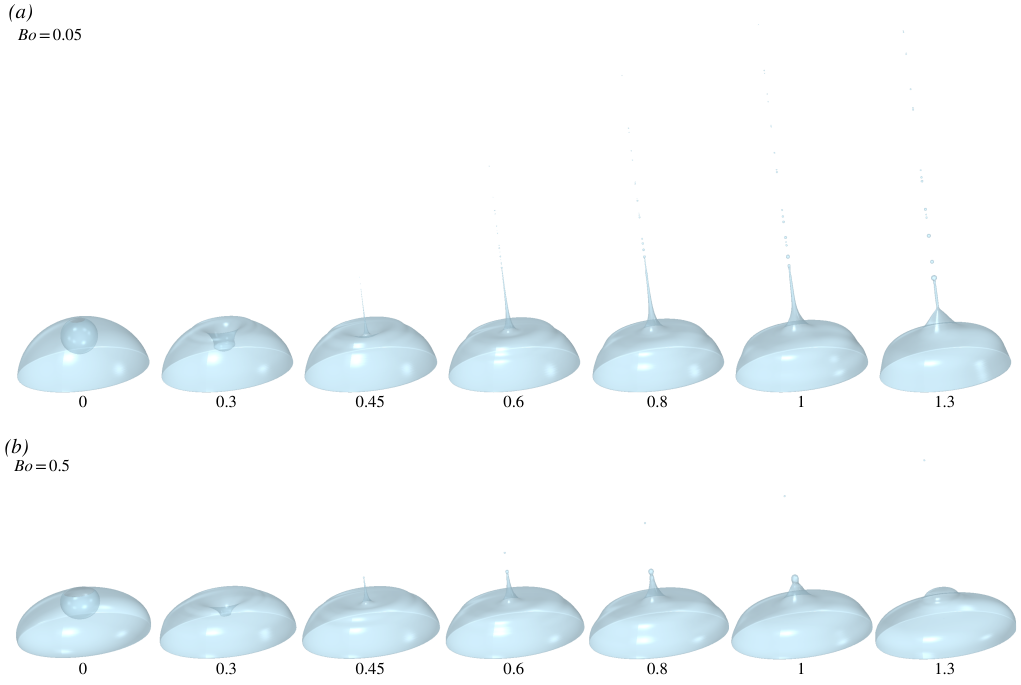}
\caption{The same drop ($R_D/R_b=3.577$, $\Lambda=16$) at $\Ohn=0.02$, $N=13$:
(\textit{a}) $\Bo=0.05$, (\textit{b}) $\Bo=0.5$, drawn as in
figure~\ref{fig:ghb} and at the same scale. Times in units of $t_\sigma$.}
\label{fig:seq}
\end{figure*}

\begin{figure*}
\centering
\includegraphics[width=\textwidth]{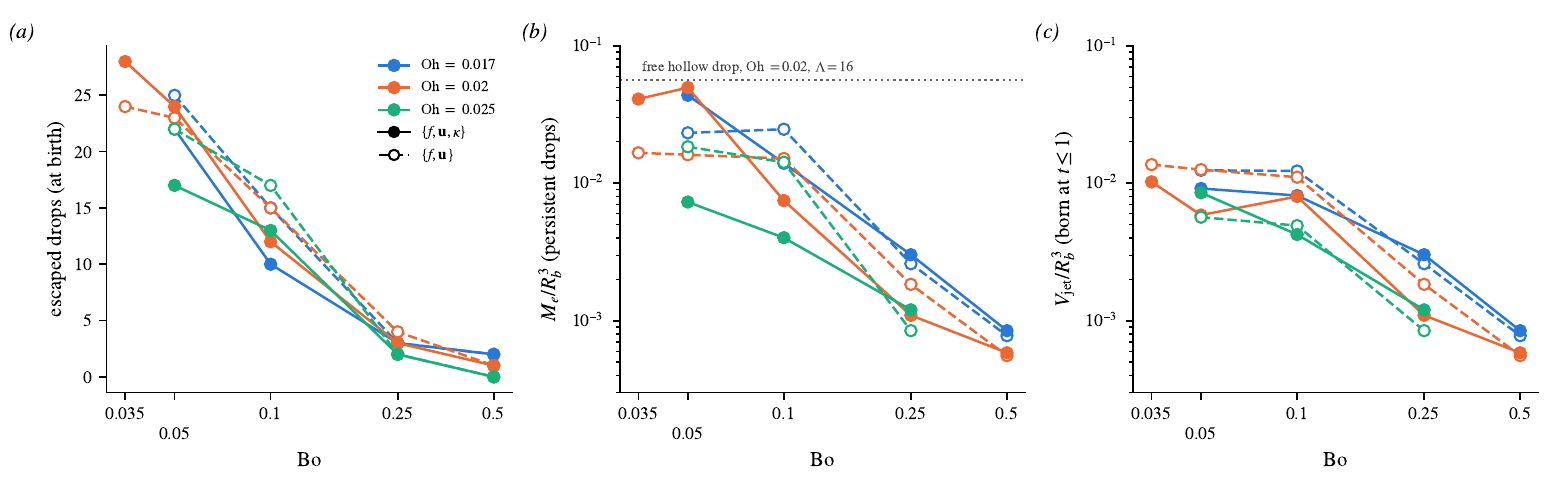}
\caption{Ejection against $\Bo$ at fixed footprint and volume, $N=13$. Filled
symbols and solid lines: criterion $\{f,\mathbf{u},\kappa\}$; open symbols and
dashed lines: $\{f,\mathbf{u}\}$. (\textit{a}) Escaped droplets counted at
birth. (\textit{b}) Ejected volume (persistent droplets); dotted line: free
hollow droplet at the same $\Ohn$ and $\Lambda$ \citep{HollowDrops2026}.
(\textit{c}) Volume of the droplets born during the jet phase, $t\le1$. At
$\Bo=0.5$, $\Ohn=0.025$ no droplet escapes.}
\label{fig:ej}
\end{figure*}

To separate gravity from geometry, the footprint and the liquid volume are held
fixed ($R_D=3.577$, $\Lambda=16$) while $\Bo$ varies from $0.035$ to $0.5$ and
$\Ohn$ from $0.017$ to $0.025$; $\Pi$ and $\alpha$ follow from
\S\,\ref{sec:eq} (table~\ref{tab:geom}), and the bubble bottom stays at
$0.95$--$1.00R_b$ above the substrate. The thirteen cases were computed at
$N=13$ with both refinement criteria. Figure~\ref{fig:seq} contrasts the two
ends of the sweep at $\Ohn=0.02$: at $\Bo=0.05$ a long, thin jet sheds a train
of about twenty droplets, while at $\Bo=0.5$ a short, thick jet sheds one.
Table~\ref{tab:census} and figure~\ref{fig:ej} give the census.

\begin{table*}
\begin{center}
\def~{\hphantom{0}}
\begin{tabular}{lccccc}
$\Bo$ & $\Delta z$ & $\Pi$ & $\alpha$ ($^\circ$) & $z_b$ & $\ell_c/R_b$\\[3pt]
0.035 & $-14.726$ & 0.515 & ~80.6 & 0.955 & 5.3\\
0.05  & $-10.061$ & 0.503 & ~81.7 & 0.965 & 4.5\\
0.1   & ~$-4.654$ & 0.465 & ~85.0 & 0.986 & 3.2\\
0.25  & ~$-1.500$ & 0.375 & ~94.5 & 1.003 & 2.0\\
0.5   & ~$-0.541$ & 0.271 & 109.0 & 0.992 & 1.4\\
\end{tabular}
\caption{Initial equilibria of the sweep ($R_D/R_b=3.577$, $\Lambda=16$):
displacement, overpressure of the drop, apparent contact angle, height of the
bubble bottom above the substrate and capillary length
$\ell_c=R_b\Bo^{-1/2}$.}
\label{tab:geom}
\end{center}
\end{table*}

\begin{table*}
\begin{center}
\def~{\hphantom{0}}
\begin{tabular}{ll|ccc|ccc}
 & & \multicolumn{3}{c|}{$\{f,\mathbf{u}\}$} & \multicolumn{3}{c}{$\{f,\mathbf{u},\kappa\}$}\\
$\Bo$ & $\Ohn$ & $n_b$ & $10^3M_e$ & $10^3V_{\rm jet}$ & $n_b$ & $10^3M_e$ & $10^3V_{\rm jet}$\\[3pt]
0.035 & 0.02  & 24 & 16.7 & 13.7 & 28 & 41.0 & 10.2\\[2pt]
0.05  & 0.017 & 25 & 23.3 & 12.4 & 22 & 43.7 & ~9.2\\
      & 0.02  & 23 & 16.1 & 12.6 & 24 & 49.6 & ~5.9\\
      & 0.025 & 22 & 18.4 & ~5.7 & 17 & ~7.3 & ~8.5\\[2pt]
0.1   & 0.017 & 15 & 24.7 & 12.3 & 10 & 14.0 & ~8.1\\
      & 0.02  & 15 & 15.1 & 11.1 & 12 & ~7.5 & ~8.0\\
      & 0.025 & 17 & 14.2 & ~4.9 & 13 & ~4.0 & ~4.2\\[2pt]
0.25  & 0.017 & ~3 & ~2.60 & ~2.60 & ~3 & ~3.02 & ~3.02\\
      & 0.02  & ~4 & ~1.84 & ~1.84 & ~3 & ~1.10 & ~1.10\\
      & 0.025 & ~2 & ~0.85 & ~0.85 & ~2 & ~1.20 & ~1.20\\[2pt]
0.5   & 0.017 & ~2 & ~0.78 & ~0.78 & ~2 & ~0.85 & ~0.85\\
      & 0.02  & ~1 & ~0.56 & ~0.56 & ~1 & ~0.58 & ~0.58\\
      & 0.025 & ~0 & 0 & 0 & ~0 & 0 & 0\\
\end{tabular}
\caption{Census of the sweep at $N=13$ with the two refinement criteria:
number of escaped droplets counted at birth, $n_b$; ejected volume (persistent
escaped droplets), $M_e$; and volume of the escaped droplets born at $t\le1$,
$V_{\rm jet}$. Volumes in units of $R_b^3$ ($V_b=4.19R_b^3$).}
\label{tab:census}
\end{center}
\end{table*}

\subsection{The jet phase: gravity sets the amount}\label{sec:jet}

The volume ejected during the jet phase, $V_{\rm jet}$, behaves in the same way
with both criteria (figure~\ref{fig:ej}\textit{c}). Averaged over $\Ohn$, it
grows by a factor of 21 (criterion $\{f,\mathbf{u}\}$) or 14
($\{f,\mathbf{u},\kappa\}$) from $\Bo=0.5$ to $\Bo=0.1$, and at $\Ohn=0.02$ by only
a further factor of $1.2$--$1.3$ from $\Bo=0.1$ to $0.035$: below
$\Bo\approx0.1$ the jet phase saturates, with $V_{\rm jet}\approx(0.6$--$1.4)\times
10^{-2}R_b^3$ at $\Ohn\le0.02$, that is $0.14$--$0.33\,\%$ of the bubble
volume. With the curvature criterion
$V_{\rm jet}$ is about 25\,\% smaller at $\Bo\le0.1$ and the same at $\Bo\ge0.25$.
The number of escaped droplets follows the same pattern (figure~\ref{fig:ej}
\textit{a}): from none to two at $\Bo=0.5$ to between 17 and 28 at $\Bo\le0.05$.

The saturation has a simple origin. Gravity enters through the capillary length
$\ell_c=R_b\Bo^{-1/2}$, which is $1.4R_b$ at $\Bo=0.5$ and grows beyond $3R_b$
for $\Bo\le0.1$ (table~\ref{tab:geom}). When $\ell_c$ is comparable to the
bubble, gravity deforms the cavity and decelerates the jet; once $\ell_c$
exceeds the cavity by a factor of a few, the cavity is spherical to leading
order (\S\,\ref{sec:eq}) and the jet no longer feels gravity during its
growth. The decrease of $\Pi$ with $\Bo$ at fixed footprint (from $0.52$ to
$0.27$, table~\ref{tab:geom}) acts in the same direction, since a smaller
overpressure focuses less energy into the jet
\citep{HollowDrops2026}.

\subsection{Late droplets}\label{sec:late}

The total ejected volume at $\Bo\le0.1$ does not saturate in the same way, and
it is the quantity that depends on the refinement criterion
(figure~\ref{fig:ej}\textit{b}). At small $\Bo$, after the jet phase, a thick
column of liquid left at the base of the jet breaks up into between none and
three large droplets ($R_d\ge0.09R_b$), born between $t\approx0.9$ and $1.9$.
Each carries a volume comparable to the whole jet phase, and their number and
size change with the criterion: with curvature refinement they nearly
disappear at $\Bo=0.1$, where $M_e$ falls to between 0.3 and 0.6 times its
value with $\{f,\mathbf{u}\}$, and they become larger at $\Bo\le0.05$, where
$M_e$ is two to three times larger in three of the four cases. Over the seven
cases with $\Bo\le0.1$ taken as realizations, the mean ejected volume is the
same within the scatter, $(1.8\pm0.4)\times10^{-2}R_b^3$ with
$\{f,\mathbf{u}\}$ and $(2.4\pm2.0)\times10^{-2}R_b^3$ with
$\{f,\mathbf{u},\kappa\}$, but the value in each case is uncertain by a factor
of two or more. The breakup of a slow, thick column is sensitive to any
perturbation, numerical or physical, and the late droplets should be regarded
as a stochastic component of the ejection. An ensemble of perturbed initial
conditions, as in \citet{Berny2021}, would be needed to give their mean. In
the inertial case of \S\,\ref{sec:GHA}, with a smaller drop and a much smaller
$\Ohn$, no late droplet escapes.

For comparison, a free hollow droplet with the same $\Lambda$ and $\Ohn=0.02$
ejects $0.056R_b^3$ \citep{HollowDrops2026}: the substrate and gravity reduce
the ejected volume by a factor of about three at $\Bo\le0.1$ (with the scatter
of the late droplets) and by thirty to fifty at $\Bo=0.25$.

\subsection{Viscosity and the emission boundary}\label{sec:visc}

The sensitivity of the jet-phase volume to viscosity grows with $\Bo$. Between
$\Ohn=0.017$ and $0.025$, $\mathrm{d}\ln V_{\rm jet}/\mathrm{d}\ln\Ohn$ is
$-2.0$, $-2.4$, $-2.9$ and $-\infty$ at $\Bo=0.05$, $0.1$, $0.25$ and $0.5$ with
$\{f,\mathbf{u}\}$, and $-0.2$, $-1.7$, $-2.4$ and $-\infty$ with
$\{f,\mathbf{u},\kappa\}$. At $\Bo=0.5$ no droplet escapes at $\Ohn=0.025$ with
either criterion (the computations reach $t=4$ without a single detachment), so
that at this footprint the emission boundary lies between $\Ohn=0.02$ and
$0.025$ for $\Bo=0.5$ and above $0.025$ for $\Bo\le0.25$: at fixed footprint,
increasing $\Bo$, which also lowers the overpressure $\Pi$
(table~\ref{tab:geom}), moves the boundary to smaller $\Ohn$. The sensitivity of the total volume $M_e$ to $\Ohn$ at
$\Bo\le0.1$ is dominated by the late droplets and is not a robust quantity.

\section{Droplet-size statistics}\label{sec:sizes}

The cases with $\Bo\le0.1$ yield enough droplets per run for a statistical
description: 141 escaped droplets counted at birth with $\{f,\mathbf{u}\}$, and
126 with $\{f,\mathbf{u},\kappa\}$. Every radius is scaled with the mean of its
own run, $\langle R_d\rangle$, which is $0.035$--$0.061R_b$ in all cases
(mean $0.047R_b$ with either criterion).

\subsection{Independence of gravity}

After scaling, the distributions at $\Bo=0.1$ and $\Bo=0.05$ cannot be told
apart (criterion $\{f,\mathbf{u}\}$, two-sample Kolmogorov--Smirnov test,
$p=0.92$), and even the unscaled radii agree ($p=0.16$). The coefficient of
variation ($0.53$) and the skewness ($1.1$--$1.6$) of $R_d/\langle R_d\rangle$
show no trend with $\Bo$. Gravity acts at the scale
of the cavity and of the jet, through $\ell_c\ge3R_b$, which is two orders of
magnitude above the droplets; it controls how much liquid the jet carries, not
how the jet fragments.

\subsection{Shape and resolution}

\begin{figure*}
\centering
\includegraphics[width=\textwidth]{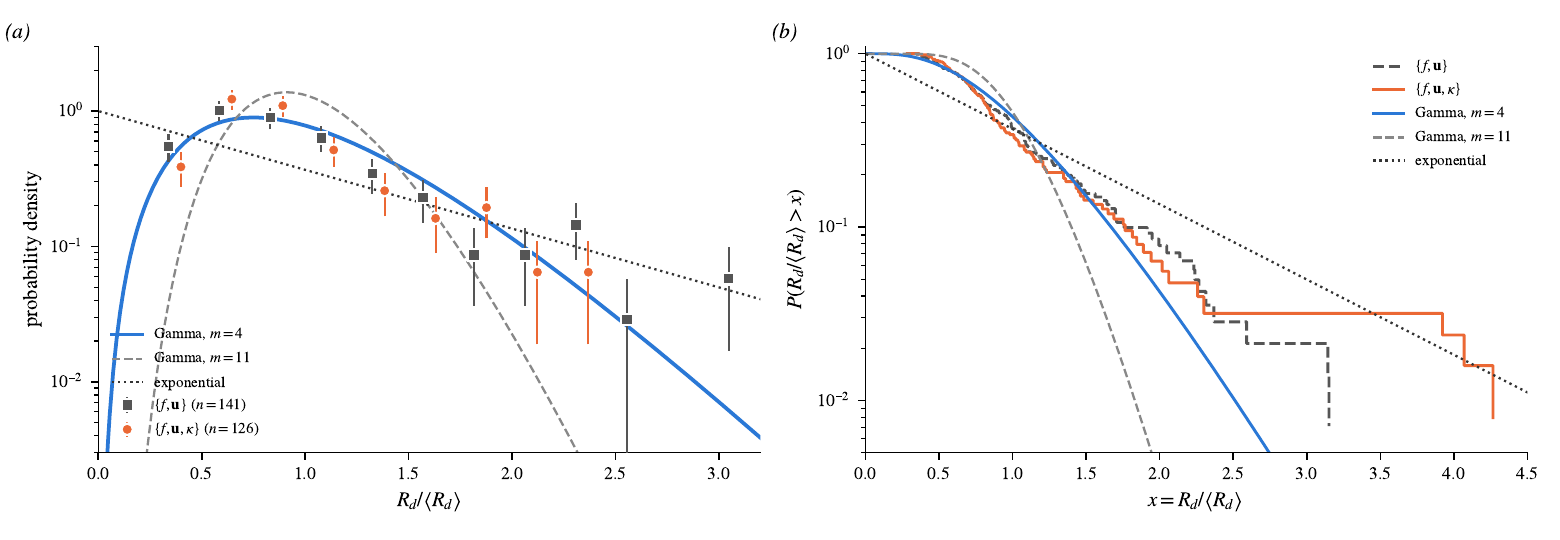}
\caption{Size distribution of the escaped droplets at birth, all cases with
$\Bo\le0.1$ pooled, each radius scaled with the mean of its run.
(\textit{a}) Probability density, with Poisson error bars. (\textit{b})
Survival function. Grey: criterion $\{f,\mathbf{u}\}$; orange:
$\{f,\mathbf{u},\kappa\}$.}
\label{fig:sizes}
\end{figure*}

Figure~\ref{fig:sizes} shows the distribution of $R_d/\langle R_d\rangle$ with
both criteria, which agree (two-sample test $p=0.59$ at birth and $p=0.89$ for
the persistent droplets). With $\sim140$ droplets per criterion the sample fixes
the tail of the distribution (\S\,\ref{sec:tail}), but not its full form. Its
left side is sharp: droplets smaller than $\approx0.02R_b$ (about $7\Delta$ at
$N=13$) are rare. In free hollow droplets the smallest droplet recorded follows
the mesh \citep{HollowDrops2026}, and the roll-off found here at a few cells is
consistent with a lower end set by the resolution; a computation at $N=14$
would confirm it in the sessile geometry. A Gamma distribution fitted to the
census has its order fixed largely by this side (maximum likelihood $3.7$--$5.1$
in the four samples, with bootstrap intervals spanning $2.8$--$6.6$), and a
log-normal distribution, or an exponential with a soft lower cut-off,
$f\propto R^{k}\exp[-(R_c/R)^3-R/\ell]$ (with $k=-0.14$,
$R_c=0.43\langle R_d\rangle$ and $\ell=0.56\langle R_d\rangle$ for
$\{f,\mathbf{u}\}$), fits it better (difference in the Akaike criterion of
$11$--$22$ at birth), although these two cannot be told apart. The order of a
fitted Gamma is therefore not a property of the sessile census that these
computations can establish, and it is not used here for comparison with the
orders reported at a bath \citep{Berny2021} or adopted in source functions
\citep{DeikeReichlPaulot2022}; that comparison requires the larger populations
of the free hollow droplet \citep{HollowDrops2026}.

\subsection{An exponential tail}\label{sec:tail}

\begin{figure*}
\centering
\includegraphics[width=\textwidth]{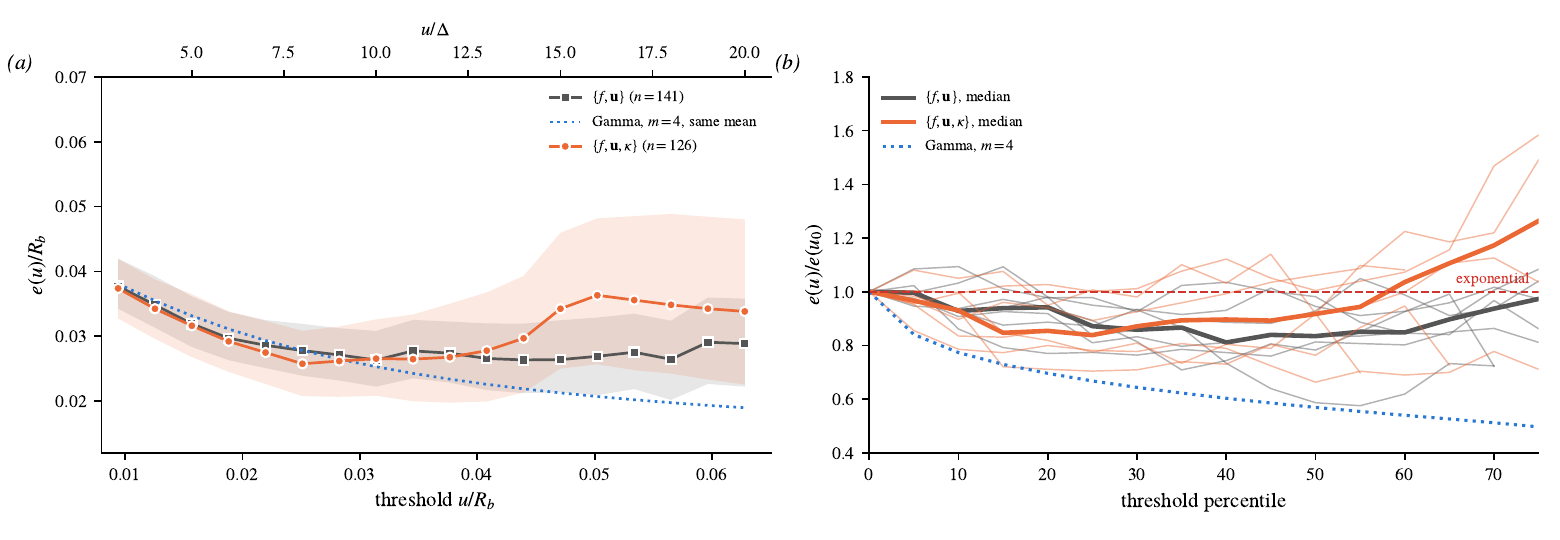}
\caption{Mean-excess function of the droplet radius, escaped droplets at
birth, $\Bo\le0.1$. (\textit{a}) All runs pooled, in absolute units from
$u=3\Delta$; bands: 90\,\% bootstrap intervals; dotted: Gamma of order four with
the same mean. (\textit{b}) Each run separately (thin lines), with thresholds at
the percentiles of its own census and $e(u)$ normalized by its value at the
smallest resolved droplet $u_0$; thick lines: medians. An exponential tail gives
$e(u)/e(u_0)=1$.}
\label{fig:mex}
\end{figure*}

The mean-excess function $e(u)=E[R_d-u\,|\,R_d\ge u]$ discriminates between
tails and is insensitive to the left side of the census: it is constant for an exponential tail and decreases for a Gamma of order
larger than one (figure~\ref{fig:mex}). With $\{f,\mathbf{u}\}$ it decreases
from $0.038R_b$ at $u=3\Delta$ to $0.026R_b$ at $u\approx0.03R_b$ and then
stays flat, the signature of an exponential tail with a scale close to
$0.6\langle R_d\rangle$, while a Gamma of order four with the same mean keeps
decreasing. With $\{f,\mathbf{u},\kappa\}$ it follows the same curve up to
$u\approx0.04R_b$ and then rises, because the larger late droplets make the
tail heavier than exponential. The test can also be made run by run, with the
thresholds placed at the percentiles of each census and $e(u)$ normalized by
its value at the smallest resolved droplet, as in \citet{HollowDrops2026}
(figure~\ref{fig:mex}\textit{b}): the median of the runs stays within 20\,\% of
the exponential value up to the 60th percentile with both criteria, while a
Gamma of order four treated in the same way falls to $0.5$.

\subsection{Coalescence in flight}\label{sec:coal}

In an axisymmetric computation every droplet travels on the axis, and a faster
droplet that catches a slower one must coalesce with it. Of the 141 droplets
that escape at birth with $\{f,\mathbf{u}\}$, only 93 persist (81 of 126 with
$\{f,\mathbf{u},\kappa\}$). In an experiment, small departures from
axisymmetry let droplets pass each other, and the birth census is the one that
should be transferred to three dimensions. High-speed imaging of the droplet
train, combined with simulations of the same configuration, could measure the
number of droplets that leave the jet, the number of coalescences in flight
and the final count; this would calibrate the axisymmetric droplet statistics
on which aerosol source functions rest.

\section{The substrate at zero Bond number}\label{sec:floor}

The limit $\Bo\to0$ of \S\,\ref{sec:eq} isolates the effect of the substrate.
At $\Bo=0$ the sessile drop is a hollow drop of radius ratio $\lambda=2/\Pi$,
a unit bubble internally tangent to a spherical drop of radius $\lambda$, cut by
a plane at height $z_b$ above the bottom of the bubble. The overpressure is fixed
by $\lambda$, and the substrate changes only the liquid below the cavity. The
free hollow drop is recovered as $z_b\to2\lambda-2$, and its ejection boundary
is $\Ohn_1=\Ohn_c(1+2\beta/\lambda)$, with $\Ohn_c=0.043$ and $\beta=0.83$
\citep{HollowDrops2026}. The term $2\beta/\lambda$ is attributed there to the
Laplace overpressure of the outer surface and to the reduced inertia of a
shell whose outer surface is free. A substrate at $\Bo=0$ leaves the first
unchanged and, by replacing part of the free surface with a rigid wall, should
remove part of the second, moving the boundary towards its flat-surface value
as it rises.

\begin{figure*}
\centering
\includegraphics[width=\textwidth]{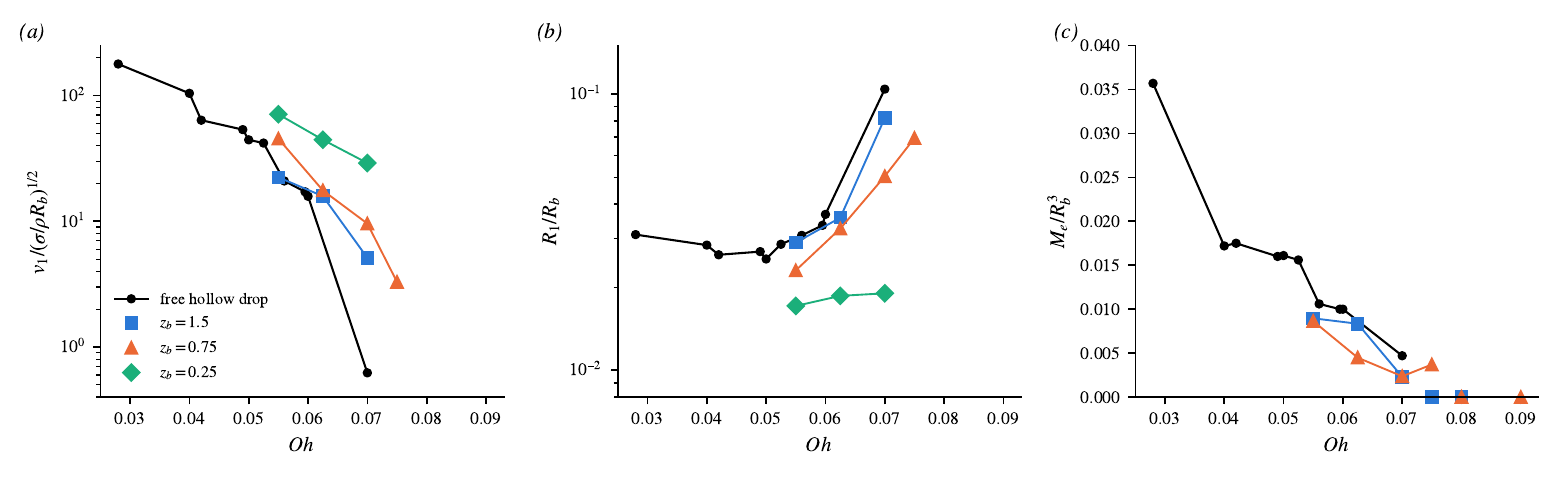}
\caption{First droplet and ejected volume for the hollow drop $\lambda=2.571$
cut by a substrate at height $z_b$ ($\Bo=0$, $N=14$), against the free hollow
drop with $\Lambda=16$ at the same resolution \citep{HollowDrops2026}.
(\textit{a}) Launch velocity and (\textit{b}) radius of the first droplet.
(\textit{c}) Ejected volume; points on the axis: no droplet escapes.}
\label{fig:floor}
\end{figure*}

We computed $\lambda=2.571$, the column $\Lambda=16$ of the free campaign, whose
boundary is $\Ohn_1=0.070\pm0.005$, with the substrate at $z_b=1.5$, $0.75$ and
$0.25$ ($\Lambda=11.9$, $8.4$ and $5.9$; $\alpha=111^\circ$, $94^\circ$ and
$83^\circ$), at $N=14$ with $\{f,\mathbf{u}\}$: first at $\Ohn=0.055$, $0.0625$
and $0.07$, and then, to locate the boundary, at $\Ohn=0.075$ and $0.08$
($z_b=1.5$) and $0.075$, $0.08$ and $0.09$ ($z_b=0.75$). As $\Bo\to0$ the film
cap shrinks to a point and the rim of \S\,\ref{sec:rim} becomes the puncture
rule of the free campaign. With $\Delta=R_b/637$ and $h_r=2.5\Delta$, the
initial condition and the resolution are those of the free computations at
their level $N=15$, which serve as the reference. Escape is measured, as in the
free campaign, from the centre of the drop sphere with the scale $\lambda$.
Table~\ref{tab:floor} and figure~\ref{fig:floor} give the first droplet, the one
leading the train, and the ejected volume.

\begin{table*}
\begin{center}
\def~{\hphantom{0}}
\begin{tabular}{lccccccc}
$z_b$ & $\Lambda$ & $\Ohn$ & $n_e$ & $10^3M_e/R_b^3$ & $R_1/R_b$ & $v_1$ & $t_1$\\[3pt]
free  & 16~~ & 0.056~ & 10 & 10.6 & 0.031 & 20.9 & 0.47\\
      &      & 0.0595 & ~8 & 10.0 & 0.034 & 17.2 & 0.49\\
      &      & 0.07~~ & ~1 & ~4.7 & 0.104 & ~0.6 & 0.94\\
      &      & 0.08~~ & ~0 & ~0~~ & --    & --   & --\\[2pt]
1.5   & 11.9 & 0.055~ & 10 & ~9.0 & 0.029 & 22.4 & 0.46\\
      &      & 0.0625 & ~6 & ~8.3 & 0.036 & 15.8 & 0.50\\
      &      & 0.07~~ & ~1 & ~2.3 & 0.082 & ~5.1 & 0.72\\
      &      & 0.075~ & ~0 & ~0~~ & --    & --   & --\\
      &      & 0.08~~ & ~0 & ~0~~ & --    & --   & --\\[2pt]
0.75  & ~8.4 & 0.055~ & 14 & ~8.6 & 0.023 & 45.4 & 0.42\\
      &      & 0.0625 & ~8 & ~4.5 & 0.033 & 17.6 & 0.50\\
      &      & 0.07~~ & ~4 & ~2.4 & 0.050 & ~9.6 & 0.58\\
      &      & 0.075~ & ~2 & ~3.7 & 0.069 & ~3.3 & 0.71\\
      &      & 0.08~~ & ~0 & ~0~~ & --    & --   & --\\
      &      & 0.09~~ & ~0 & ~0~~ & --    & --   & --\\[2pt]
0.25  & ~5.9 & 0.055~ & -- & -- & 0.017 & 71.0 & 0.43\\
      &      & 0.0625 & -- & -- & 0.019 & 44.5 & 0.45\\
      &      & 0.07~~ & -- & -- & 0.019 & 29.1 & 0.47\\
\end{tabular}
\caption{Hollow drop $\lambda=2.571$ cut by a substrate at height $z_b$ above
the bottom of the bubble, $\Bo=0$, $N=14$: number $n_e$ and volume $M_e$ of the
escaped droplets, and radius $R_1$, launch velocity $v_1$ (in units of
$(\sigma/\rho R_b)^{1/2}$) and time of birth $t_1$ of the first droplet. Free:
free hollow drop with $\Lambda=16$ at the same resolution
\citep{HollowDrops2026}, averaged over three realizations at $\Ohn=0.056$ and
$0.0595$. At $z_b=0.25$, where the computations are much slower, only the first
droplet is reported.}
\label{tab:floor}
\end{center}
\end{table*}

The result is the opposite of the expectation. At every $\Ohn$, raising the
substrate makes the first droplet faster, smaller and earlier and increases the
number of escaped droplets: the bursting moves away from its ejection boundary.
At $\Ohn=0.07$, where the free drop sits on its boundary (one droplet of radius
$0.104$, launched at $v_1=0.6$ at $t=0.94$), the substrate at $z_b=1.5$ gives one
droplet at $v_1=5.1$, at $z_b=0.75$ four droplets led by one at $v_1=9.6$, and at
$z_b=0.25$ a train led by a droplet at $v_1=29$.

The boundary itself was located by bisection. At $z_b=1.5$ droplets escape at
$\Ohn=0.07$ but not at $0.075$ or $0.08$: the boundary lies between $0.070$ and
$0.075$, indistinguishable from that of the free drop. At $z_b=0.75$ two
droplets escape at $\Ohn=0.075$, led by one at $v_1=3.3$, and none at $0.08$ or
$0.09$: the boundary lies between $0.075$ and $0.08$, 7 to 14\,\% above the free
one. In both cases the fragments that detach above the boundary fall back into
the drop. The same result follows from the first droplet alone. Since $v_1$
falls steeply and monotonically towards the boundary of the free drop, each
case can be assigned the Ohnesorge number $\Ohn_{\rm eq}$ at which the free drop
launches its first droplet at the same speed; if the substrate displaces the
boundary without changing how the first droplet approaches it, the boundary is
$\Ohn^*=\Ohn_1\Ohn/\Ohn_{\rm eq}$. Over $\Ohn=0.055$--$0.07$ the ratio
$\Ohn/\Ohn_{\rm eq}$ is $1.02\pm0.03$ at $z_b=1.5$ and $1.08\pm0.02$ at
$z_b=0.75$, so that $\Ohn^*\simeq0.071$ and $0.076$, both inside the brackets
found by bisection. At $z_b=0.25$ the same matching gives $1.28\pm0.04$,
$\Ohn^*\simeq0.09$, but there the premise fails: the first droplet is about
a third smaller than the smallest first droplet of the free drop
(figure~\ref{fig:floor}\textit{b}), the jet itself has changed, and the value is
only indicative. The substrate thus raises the ejection boundary:
imperceptibly when it lies one and a half bubble radii below the bubble, by
about a tenth at $z_b=0.75$ and, on the indicative value at $z_b=0.25$, by more
as it approaches the cavity.

The ejected volume moves the other way. At $\Ohn=0.0625$ the free drop would
eject about $9\times10^{-3}R_b^3$ (interpolating table~\ref{tab:floor}); the
substrate at $z_b=1.5$ ejects the same, and at $z_b=0.75$ about half, in more
droplets. At $\Ohn=0.07$ both substrates eject half of the single droplet of
the free drop. At the same Ohnesorge number the substrate gives faster
droplets and, close to the cavity, smaller ones; at the same distance from the
boundary, measured by $\Ohn_{\rm eq}$, it ejects less liquid: at $\Ohn=0.0625$,
$\Ohn_{\rm eq}\simeq0.059$--$0.060$, where the free drop ejects
$1.0\times10^{-2}R_b^3$, against $0.83$ and $0.45\times10^{-2}R_b^3$ with the
substrate at $z_b=1.5$ and $0.75$. Right at the boundary the volume is
carried by a few large, slow droplets: at $z_b=0.75$ the two ejected at
$\Ohn=0.075$ carry more ($3.7\times10^{-3}R_b^3$) than the four ejected at
$0.07$.

Neither trend fits the inertia argument. We interpret them through the flow
below the cavity. The collapse of the cavity drives liquid both upwards, into
the jet, and downwards, away from the bottom of the cavity. A rigid wall a
distance $z_b$ below the bubble blocks the downward flow and turns it back
towards the axis and upwards: its image reinforces the convergence at the
bottom of the cavity, which produces a faster and thinner jet that carries
less liquid. The effect grows rapidly as the wall approaches the cavity, and at
$z_b=0.25$ it changes the jet. In the sessile drops of \S\,\ref{sec:sweep},
increasing $\Bo$ at fixed footprint, together with the decrease of $\Pi$ that it
brings, moves the boundary the other way, to smaller $\Ohn$
(\S\,\ref{sec:visc}); the ejection threshold of a sessile drop results from the
competition between the substrate on the one hand and gravity and overpressure
on the other.

\section{Computed and measured ejected volume}\label{sec:Ve}

\citet{GutierrezEtal2026} measured the total volume emitted by the bursting,
$V_e$, as the difference between the volume of the drop before the injection of
the bubble and after the ejection, both obtained from side images by
sub-pixel detection of the contour. They report $\langle V_e\rangle\simeq
0.054R_D^3$, independent of the confinement within their range, and a first
droplet of about $1\,\%$ of $V_e$. In the geometry of \S\,\ref{sec:GHA}
($R_D=1.99R_b$) this is $V_e\simeq0.43R_b^3$, or 10\,\% of the bubble volume.

The simulations of \S\,\ref{sec:GHA} eject $0.044$--$0.053R_b^3$ with the four
set-ups, and nothing after $t=1.1$ in the two runs continued to the automatic
stop. The volume lost by the mother drop, which is the computational
counterpart of the measured difference, is $0.061R_b^3$ in the run with
curvature refinement, in which the total liquid volume is conserved to
$10^{-5}$. The largest volume detached at any instant, including the fragments
that later fall back, is $0.077R_b^3$. The measured volume is therefore eight
to ten times what the simulations eject, seven times what the mother drop
loses and more than five times the largest detached volume, and none of the choices examined
here (initial condition, refinement criterion, duration of the computation)
reduces the difference. The late droplets of \S\,\ref{sec:late}, which carry
most of the volume in the sweep at small $\Bo$, do not appear in this geometry.

Since $V_e$ is less than 1\,\% of the volume of the drop, it is a small difference
between two large volumes measured from images, and several contributions
outside the jet could enter it. Two can be estimated. A millimetric sessile
water drop evaporates by diffusion at a rate of order
$\pi R_D\mathcal{D}c_s(1-\phi)f(\alpha)$ \citep{Popov2005}, with $\mathcal{D}$
the diffusivity of vapour in air, $c_s$ the saturation concentration, $\phi$
the relative humidity and $f(\alpha)$ a function of order 2--4 for the contact
angles involved; at room temperature and $\phi=0.5$ this is
$\sim5\times10^{-3}\,\mathrm{mm^3\,s^{-1}}$, so that the measured $V_e$
($\approx0.43\,\mathrm{mm}^3$ for $R_b\approx1\,\mathrm{mm}$) would be lost in
about a minute and a half. Evaporation is therefore a possible contribution,
and the conditions that would allow it to be excluded (the time between the two
images, the humidity, any enclosure of the drop) are not reported. Liquid drawn into the orifice through which
the bubble is injected would also count as emitted. A like-for-like comparison
requires knowing what $V_e$ contains, and the simulations give a bound on the
part due to the jet.

\section{Conclusions}\label{sec:concl}

The sessile drop is a convenient laboratory for confined bubble bursting, and
it can be computed without arbitrary choices. Its initial equilibria form a
single family obtained from the bath by a vertical displacement, whose
natural measure is the overpressure of the drop, $\Pi=-\Bo\,\Delta z$, and whose
small-Bond-number limit, singular at a bath, is regular in a drop, where it
reduces to a hollow drop cut by the substrate. The film edge can be represented by a
Taylor--Culick rim built from the equilibrium, with the volume of the wedge it
replaces and the momentum of the retracting film, which removes the free
regularization parameter of previous computations. With this initial
condition, axisymmetric simulations eject droplets below the bath threshold,
as observed experimentally, and reproduce the measured velocity of the first
droplet within 7\,\%; the measured radius of the first droplet lies between the values
obtained with the two refinement criteria.

At fixed drop footprint and volume, gravity controls how much liquid the jet
ejects, not how it fragments. The volume ejected during the jet phase grows by
a factor of 14--21 from $\Bo=0.5$ to $0.1$ and saturates below, where the
capillary length exceeds the cavity by a factor of three or more; the
sensitivity to viscosity grows with $\Bo$ and places the emission boundary at
smaller $\Ohn$ the larger $\Bo$, which at fixed footprint also lowers the
overpressure. At small $\Bo$ a few large, late droplets add a stochastic
component to the total ejected volume, uncertain by a factor of two or more in
each realization; its mean requires an ensemble of perturbed computations.

The size distribution of the jet droplets, scaled with its mean, is
independent of $\Bo$ and of the refinement criterion, unimodal, with an
exponential tail; its lower end, at a few cells, is consistent with a limit set
by the resolution. Since the free hollow droplet
gives the same exponential census \citep{HollowDrops2026}, the shape of the
census appears to be a property of jet fragmentation that confinement and
gravity leave unchanged.

At zero Bond number, where the sessile drop is a hollow droplet cut by the
substrate, the substrate alone raises the ejection boundary, imperceptibly
when it lies well below the bubble, by about a tenth at three quarters of a
bubble radius and increasingly as it approaches the cavity, while at the same
distance from the boundary the bursting ejects less liquid. The flow beneath
the cavity, blocked and turned upwards by the wall, offers an interpretation.
Gravity, with the decrease of the overpressure that it brings at fixed
footprint, acts in the opposite direction, so that the ejection threshold of a
sessile drop results from the competition between them.

Finally, the total volume emitted in experiments exceeds the volume the simulations eject in the same geometry by a factor of
eight to ten. Identifying what the measured volume contains, and measuring
the droplet train and its coalescences in flight with high-speed imaging, are
the two experiments that would most sharpen the comparison between computed
and measured bursting statistics.

\paragraph*{Supplementary material.}
A table of sessile equilibria ($10^{-4}\le\Bo\le70$, $0\le\Pi\le2$) and the initial-condition generator are available as
supplementary material.

\section*{Acknowledgements}
The author thanks D. Fern\'andez Rivas and U. J. Guti\'errez-Hern\'andez for
discussions in September 2024 in which the sessile configuration was first
considered. Jos\'e M. L\'opez-Herrera and Miguel A. Herrada provided highly
appreciated computational support.

\section*{Use of AI tools}
The author used a large language model (Claude, Anthropic) as an assistant for
post-processing the simulation output, for statistical fitting, for preparing
figures and for editing the text of the manuscript. The study was designed and
conducted by the author, who verified all results and takes full
responsibility for the content.

\section*{Funding}
This work was partially supported by grant PID2022-140951OB-C21, funded by
MCIN/\allowbreak AEI/\allowbreak 10.13039/\allowbreak 501100011033 and by ERDF/EU, within Line 1 (\textit{Ultrafine
aerosols and their generation from surface bubble bursting}) of the coordinated project PID2022-140951OB-C21/\allowbreak C22. 

\section*{Declaration of interests}
The author reports no conflict of interest.

\section*{Data availability statement}
The data that support the findings of this study are available from the author
upon reasonable request.

\bibliographystyle{apsrev4-2}
\bibliography{sessile}

\begin{thebibliography}{18}%
\makeatletter
\providecommand \@ifxundefined [1]{%
 \@ifx{#1\undefined}
}%
\providecommand \@ifnum [1]{%
 \ifnum #1\expandafter \@firstoftwo
 \else \expandafter \@secondoftwo
 \fi
}%
\providecommand \@ifx [1]{%
 \ifx #1\expandafter \@firstoftwo
 \else \expandafter \@secondoftwo
 \fi
}%
\providecommand \natexlab [1]{#1}%
\providecommand \enquote  [1]{``#1''}%
\providecommand \bibnamefont  [1]{#1}%
\providecommand \bibfnamefont [1]{#1}%
\providecommand \citenamefont [1]{#1}%
\providecommand \href@noop [0]{\@secondoftwo}%
\providecommand \href [0]{\begingroup \@sanitize@url \@href}%
\providecommand \@href[1]{\@@startlink{#1}\@@href}%
\providecommand \@@href[1]{\endgroup#1\@@endlink}%
\providecommand \@sanitize@url [0]{\catcode `\\12\catcode `\$12\catcode
  `\&12\catcode `\#12\catcode `\^12\catcode `\_12\catcode `\%12\relax}%
\providecommand \@@startlink[1]{}%
\providecommand \@@endlink[0]{}%
\providecommand \url  [0]{\begingroup\@sanitize@url \@url }%
\providecommand \@url [1]{\endgroup\@href {#1}{\urlprefix }}%
\providecommand \urlprefix  [0]{URL }%
\providecommand \Eprint [0]{\href }%
\providecommand \doibase [0]{https://doi.org/}%
\providecommand \selectlanguage [0]{\@gobble}%
\providecommand \bibinfo  [0]{\@secondoftwo}%
\providecommand \bibfield  [0]{\@secondoftwo}%
\providecommand \translation [1]{[#1]}%
\providecommand \BibitemOpen [0]{}%
\providecommand \bibitemStop [0]{}%
\providecommand \bibitemNoStop [0]{.\EOS\space}%
\providecommand \EOS [0]{\spacefactor3000\relax}%
\providecommand \BibitemShut  [1]{\csname bibitem#1\endcsname}%
\let\auto@bib@innerbib\@empty
\bibitem [{\citenamefont {Spiel}(1994)}]{Spiel1994}%
  \BibitemOpen
  \bibfield  {author} {\bibinfo {author} {\bibfnamefont {D.~E.}\ \bibnamefont
  {Spiel}},\ }\href {https://doi.org/10.1029/94JC00382} {\bibfield  {journal}
  {\bibinfo  {journal} {J. Geophys. Res. Oceans}\ }\textbf {\bibinfo {volume}
  {99}},\ \bibinfo {pages} {10289} (\bibinfo {year} {1994})}\BibitemShut
  {NoStop}%
\bibitem [{\citenamefont {Duchemin}\ \emph {et~al.}(2002)\citenamefont
  {Duchemin}, \citenamefont {Popinet}, \citenamefont {Josserand},\ and\
  \citenamefont {Zaleski}}]{DuchEtal2002}%
  \BibitemOpen
  \bibfield  {author} {\bibinfo {author} {\bibfnamefont {L.}~\bibnamefont
  {Duchemin}}, \bibinfo {author} {\bibfnamefont {S.}~\bibnamefont {Popinet}},
  \bibinfo {author} {\bibfnamefont {C.}~\bibnamefont {Josserand}},\ and\
  \bibinfo {author} {\bibfnamefont {S.}~\bibnamefont {Zaleski}},\ }\href@noop
  {} {\bibfield  {journal} {\bibinfo  {journal} {Phys. Fluids}\ }\textbf
  {\bibinfo {volume} {14}},\ \bibinfo {pages} {3000} (\bibinfo {year}
  {2002})}\BibitemShut {NoStop}%
\bibitem [{\citenamefont {Lhuissier}\ and\ \citenamefont
  {Villermaux}(2012)}]{LhuiVill2012}%
  \BibitemOpen
  \bibfield  {author} {\bibinfo {author} {\bibfnamefont {H.}~\bibnamefont
  {Lhuissier}}\ and\ \bibinfo {author} {\bibfnamefont {E.}~\bibnamefont
  {Villermaux}},\ }\href@noop {} {\bibfield  {journal} {\bibinfo  {journal} {J.
  Fluid Mech.}\ }\textbf {\bibinfo {volume} {696}},\ \bibinfo {pages} {5}
  (\bibinfo {year} {2012})}\BibitemShut {NoStop}%
\bibitem [{\citenamefont {Veron}(2015)}]{Veron2015}%
  \BibitemOpen
  \bibfield  {author} {\bibinfo {author} {\bibfnamefont {F.}~\bibnamefont
  {Veron}},\ }\href@noop {} {\bibfield  {journal} {\bibinfo  {journal} {Annu.
  Rev. Fluid Mech.}\ }\textbf {\bibinfo {volume} {47}},\ \bibinfo {pages} {507}
  (\bibinfo {year} {2015})}\BibitemShut {NoStop}%
\bibitem [{\citenamefont {Ga{\~n}{\'a}n-Calvo}(2017)}]{GananCalvo2017}%
  \BibitemOpen
  \bibfield  {author} {\bibinfo {author} {\bibfnamefont {A.~M.}\ \bibnamefont
  {Ga{\~n}{\'a}n-Calvo}},\ }\href@noop {} {\bibfield  {journal} {\bibinfo
  {journal} {Phys. Rev. Lett.}\ }\textbf {\bibinfo {volume} {119}},\ \bibinfo
  {pages} {204502} (\bibinfo {year} {2017})}\BibitemShut {NoStop}%
\bibitem [{\citenamefont {Deike}\ \emph {et~al.}(2018)\citenamefont {Deike},
  \citenamefont {Ghabache}, \citenamefont {Liger-Belair}, \citenamefont {Das},
  \citenamefont {Zaleski}, \citenamefont {Popinet},\ and\ \citenamefont
  {S{\'e}on}}]{Deike2018}%
  \BibitemOpen
  \bibfield  {author} {\bibinfo {author} {\bibfnamefont {L.}~\bibnamefont
  {Deike}}, \bibinfo {author} {\bibfnamefont {E.}~\bibnamefont {Ghabache}},
  \bibinfo {author} {\bibfnamefont {G.}~\bibnamefont {Liger-Belair}}, \bibinfo
  {author} {\bibfnamefont {A.~K.}\ \bibnamefont {Das}}, \bibinfo {author}
  {\bibfnamefont {S.}~\bibnamefont {Zaleski}}, \bibinfo {author} {\bibfnamefont
  {S.}~\bibnamefont {Popinet}},\ and\ \bibinfo {author} {\bibfnamefont
  {T.}~\bibnamefont {S{\'e}on}},\ }\href
  {https://doi.org/10.1103/PhysRevFluids.3.013603} {\bibfield  {journal}
  {\bibinfo  {journal} {Phys. Rev. Fluids}\ }\textbf {\bibinfo {volume} {3}},\
  \bibinfo {pages} {013603} (\bibinfo {year} {2018})}\BibitemShut {NoStop}%
\bibitem [{\citenamefont {Deike}\ \emph {et~al.}(2022)\citenamefont {Deike},
  \citenamefont {Reichl},\ and\ \citenamefont
  {Paulot}}]{DeikeReichlPaulot2022}%
  \BibitemOpen
  \bibfield  {author} {\bibinfo {author} {\bibfnamefont {L.}~\bibnamefont
  {Deike}}, \bibinfo {author} {\bibfnamefont {B.~G.}\ \bibnamefont {Reichl}},\
  and\ \bibinfo {author} {\bibfnamefont {F.}~\bibnamefont {Paulot}},\ }\href
  {https://doi.org/10.1029/2022AV000750} {\bibfield  {journal} {\bibinfo
  {journal} {AGU Advances}\ }\textbf {\bibinfo {volume} {3}},\ \bibinfo {pages}
  {e2022AV000750} (\bibinfo {year} {2022})}\BibitemShut {NoStop}%
\bibitem [{\citenamefont {Ga{\~n}{\'a}n-Calvo}(2026)}]{HollowDrops2026}%
  \BibitemOpen
  \bibfield  {author} {\bibinfo {author} {\bibfnamefont {A.~M.}\ \bibnamefont
  {Ga{\~n}{\'a}n-Calvo}},\ }\href@noop {} {\bibinfo {title} {The bursting of
  hollow droplets}} (\bibinfo {year} {2026}),\ \bibinfo {note}
  {arXiv:2608.26296}\BibitemShut {NoStop}%
\bibitem [{\citenamefont {Guti{\'e}rrez-Hern{\'a}ndez}\ \emph
  {et~al.}(2026)\citenamefont {Guti{\'e}rrez-Hern{\'a}ndez}, \citenamefont
  {Mu{\~n}oz-S{\'a}nchez}, \citenamefont {Ag{\'u}ndez-Cruz}, \citenamefont
  {Montanero}, \citenamefont {Fern{\'a}ndez~Rivas},\ and\ \citenamefont
  {Cabezas}}]{GutierrezEtal2026}%
  \BibitemOpen
  \bibfield  {author} {\bibinfo {author} {\bibfnamefont {U.~J.}\ \bibnamefont
  {Guti{\'e}rrez-Hern{\'a}ndez}}, \bibinfo {author} {\bibfnamefont
  {B.}~\bibnamefont {Mu{\~n}oz-S{\'a}nchez}}, \bibinfo {author} {\bibfnamefont
  {A.}~\bibnamefont {Ag{\'u}ndez-Cruz}}, \bibinfo {author} {\bibfnamefont
  {J.~M.}\ \bibnamefont {Montanero}}, \bibinfo {author} {\bibfnamefont
  {D.}~\bibnamefont {Fern{\'a}ndez~Rivas}},\ and\ \bibinfo {author}
  {\bibfnamefont {M.~G.}\ \bibnamefont {Cabezas}},\ }\href@noop {} {\bibfield
  {journal} {\bibinfo  {journal} {arXiv preprint arXiv:2607.22207}\ } (\bibinfo
  {year} {2026})}\BibitemShut {NoStop}%
\bibitem [{\citenamefont {Taylor}(1959)}]{Taylor1959}%
  \BibitemOpen
  \bibfield  {author} {\bibinfo {author} {\bibfnamefont {G.~I.}\ \bibnamefont
  {Taylor}},\ }\href@noop {} {\bibfield  {journal} {\bibinfo  {journal} {Proc.
  R. Soc. Lond. A}\ }\textbf {\bibinfo {volume} {253}},\ \bibinfo {pages} {313}
  (\bibinfo {year} {1959})}\BibitemShut {NoStop}%
\bibitem [{\citenamefont {Culick}(1960)}]{Culick1960}%
  \BibitemOpen
  \bibfield  {author} {\bibinfo {author} {\bibfnamefont {F.~E.~C.}\
  \bibnamefont {Culick}},\ }\href@noop {} {\bibfield  {journal} {\bibinfo
  {journal} {J. Appl. Phys.}\ }\textbf {\bibinfo {volume} {31}},\ \bibinfo
  {pages} {1128} (\bibinfo {year} {1960})}\BibitemShut {NoStop}%
\bibitem [{\citenamefont {Berny}\ \emph {et~al.}(2021)\citenamefont {Berny},
  \citenamefont {Popinet}, \citenamefont {S\'eon},\ and\ \citenamefont
  {Deike}}]{Berny2021}%
  \BibitemOpen
  \bibfield  {author} {\bibinfo {author} {\bibfnamefont {A.}~\bibnamefont
  {Berny}}, \bibinfo {author} {\bibfnamefont {S.}~\bibnamefont {Popinet}},
  \bibinfo {author} {\bibfnamefont {T.}~\bibnamefont {S\'eon}},\ and\ \bibinfo
  {author} {\bibfnamefont {L.}~\bibnamefont {Deike}},\ }\href
  {https://doi.org/10.1029/2021GL092919} {\bibfield  {journal} {\bibinfo
  {journal} {Geophys. Res. Lett.}\ }\textbf {\bibinfo {volume} {48}},\ \bibinfo
  {pages} {e2021GL092919} (\bibinfo {year} {2021})}\BibitemShut {NoStop}%
\bibitem [{\citenamefont {Toba}(1959)}]{Toba1959}%
  \BibitemOpen
  \bibfield  {author} {\bibinfo {author} {\bibfnamefont {Y.}~\bibnamefont
  {Toba}},\ }\href@noop {} {\bibfield  {journal} {\bibinfo  {journal} {J.
  Oceanogr. Soc. Japan}\ }\textbf {\bibinfo {volume} {15}},\ \bibinfo {pages}
  {121} (\bibinfo {year} {1959})}\BibitemShut {NoStop}%
\bibitem [{\citenamefont {Howell}(1999)}]{Howell1999}%
  \BibitemOpen
  \bibfield  {author} {\bibinfo {author} {\bibfnamefont {P.~D.}\ \bibnamefont
  {Howell}},\ }\href@noop {} {\bibfield  {journal} {\bibinfo  {journal} {J.
  Engng Maths}\ }\textbf {\bibinfo {volume} {35}},\ \bibinfo {pages} {251}
  (\bibinfo {year} {1999})}\BibitemShut {NoStop}%
\bibitem [{\citenamefont {Poulain}\ \emph {et~al.}(2018)\citenamefont
  {Poulain}, \citenamefont {Villermaux},\ and\ \citenamefont
  {Bourouiba}}]{PoulainEtal2018}%
  \BibitemOpen
  \bibfield  {author} {\bibinfo {author} {\bibfnamefont {S.}~\bibnamefont
  {Poulain}}, \bibinfo {author} {\bibfnamefont {E.}~\bibnamefont
  {Villermaux}},\ and\ \bibinfo {author} {\bibfnamefont {L.}~\bibnamefont
  {Bourouiba}},\ }\href@noop {} {\bibfield  {journal} {\bibinfo  {journal} {J.
  Fluid Mech.}\ }\textbf {\bibinfo {volume} {851}},\ \bibinfo {pages} {636}
  (\bibinfo {year} {2018})}\BibitemShut {NoStop}%
\bibitem [{\citenamefont {Popinet}(2009)}]{Popinet2009}%
  \BibitemOpen
  \bibfield  {author} {\bibinfo {author} {\bibfnamefont {S.}~\bibnamefont
  {Popinet}},\ }\href@noop {} {\bibfield  {journal} {\bibinfo  {journal} {J.
  Comput. Phys.}\ }\textbf {\bibinfo {volume} {228}},\ \bibinfo {pages} {5838}
  (\bibinfo {year} {2009})}\BibitemShut {NoStop}%
\bibitem [{\citenamefont {Walls}\ \emph {et~al.}(2015)\citenamefont {Walls},
  \citenamefont {Henaux},\ and\ \citenamefont {Bird}}]{WallsEtal2015}%
  \BibitemOpen
  \bibfield  {author} {\bibinfo {author} {\bibfnamefont {P.~L.~L.}\
  \bibnamefont {Walls}}, \bibinfo {author} {\bibfnamefont {L.}~\bibnamefont
  {Henaux}},\ and\ \bibinfo {author} {\bibfnamefont {J.~C.}\ \bibnamefont
  {Bird}},\ }\href@noop {} {\bibfield  {journal} {\bibinfo  {journal} {Phys.
  Rev. E}\ }\textbf {\bibinfo {volume} {92}},\ \bibinfo {pages} {021002(R)}
  (\bibinfo {year} {2015})}\BibitemShut {NoStop}%
\bibitem [{\citenamefont {Popov}(2005)}]{Popov2005}%
  \BibitemOpen
  \bibfield  {author} {\bibinfo {author} {\bibfnamefont {Y.~O.}\ \bibnamefont
  {Popov}},\ }\href@noop {} {\bibfield  {journal} {\bibinfo  {journal} {Phys.
  Rev. E}\ }\textbf {\bibinfo {volume} {71}},\ \bibinfo {pages} {036313}
  (\bibinfo {year} {2005})}\BibitemShut {NoStop}%
\end{thebibliography}%

\end{document}